\documentclass[%
 aip,
rsi,%
 amsmath,amssymb,
 preprint,%
]{revtex4-1}

\usepackage{siunitx}
\usepackage{graphicx}
\usepackage{dcolumn}% Align table columns on decimal point
\usepackage{bm}% bold math

\usepackage[caption=false]{subfig} % and apparently subcaption doesn't work with revtex

\begin{document}
\preprint{}

%\title{Dispersion compensation of entangled photons in standard telecommunications fiber}
\title{Characterizing nonlocal dispersion compensation in deployed telecommunications fiber}

\author{James A. Grieve}
\email{james.grieve@nus.edu.sg}
\author{Yicheng Shi}
\author{Hou Shun Poh}
\affiliation{Centre for Quantum Technologies, 3 Science Drive 2, National University of Singapore, 117543 Singapore}

\author{Christian Kurtsiefer}
\author{Alexander Ling}
\affiliation{Centre for Quantum Technologies, 3 Science Drive 2, National University of Singapore, 117543 Singapore}
\affiliation{Department of Physics, National University of Singapore, Blk S12, 2 Science Drive 3, 117551 Singapore}

\date{\today}

% To be edited by editor
%\dates{Compiled \today}

%\ociscodes{(060.5530) Pulse propagation and temporal solitons; (270.0270) Quantum optics; (060.5565) Quantum communications.}

% 060.5530   Pulse propagation and temporal solitons
% 190.4410   Nonlinear optics, parametric processes
% 060.5565   Quantum communications
% 190.4180   Multiphoton processes
% 270.0270   Quantum optics

% To be edited by editor
% \doi{\url{http://dx.doi.org/10.1364/optica.XX.XXXXXX}}

\begin{abstract}
Propagation of broadband photon pairs over deployed telecommunication fibers is used to achieve nonlocal dispersion compensation without the deliberate introduction of negative dispersion. This is made possible by exploiting time-energy entanglement and the positive and negative dispersive properties of the fiber. We demonstrate preservation of photon timing correlations after transmission over two multi-segment \SI{10}{\kilo\meter} spans of deployed fiber and up to \SI{80}{\kilo\meter} of laboratory-based fiber.
\end{abstract}

% PACS codes
% 42.50.−p Quantum Optics
% 42.65.Tg Optical solitons; nonlinear guided waves
% 42.65.Lm Parametric down conversion and production of entangled photons
% 42.79.Sz Optical communication systems, multiplexers, and demultiplexers
%\pacs{42.50.−p, 42.79.Sz, 42.65.Lm}

\maketitle

%The management of dispersion in fiber optic communications is a routine task~\cite{Keiser2003}, with compensation techniques and engineering of dispersion~\cite{Ainslie1986} enjoying long histories. While classical telecommunication tasks may be accomplished using narrow bandwidth lasers, the emerging field of quantum communications often requires the use of photon pairs created via spontaneous parametric downconversion (SPDC), an inherently broadband process.

Correlated photon pairs created via spontaneous parametric downconversion (SPDC) are a core component in entanglement based quantum key distribution (QKD)~\cite{Jennewein2000,Ribordy2001,Poppe2004,Ursin2007,Hubel2007,Marcikic2006,Honjo2008,Peloso2009,Inagaki2013,Yin2017}, and may also be used as a resource for clock synchronization~\cite{Valencia2004,Ho2009}. Photon pairs produced by this mechanism are created within a short time window (typically \SIrange{10}{100}{\femto\second}~\cite{ODonnell2011}), and so share a high degree of temporal correlation. As the SPDC process is inherently broadband, fiber chromatic dispersion can obscure these timing correlations. For this reason, photon pair sources are often filtered spectrally prior to use in optical fiber systems, reducing the throughput of the entire system~\cite{Fasel2004, Wengerowsky2018}.

Management and engineering of dispersion are routine tasks in fiber optic communications~\cite{Keiser2003, Ainslie1986}. In 1992, Franson~\cite{Franson1992} showed that photon pairs entangled in the time-energy basis could experience nonlocal compensation of chromatic dispersion, provided the photons propagate through media with opposite dispersion coefficients. This is a direct consequence of quantum correlations, and therefore impossible to replicate with classical light -- a concept later expanded by Wasak~et~al.~\cite{Wasak2010}, who proposed the preservation of tight timing correlations in the presence of dispersive transmission as an entanglement witness.

%The mechanism of nonlocal dispersion compensation is understood by considering the interplay between the tight energy anticorrelation of an entangled photon pair and the action of dispersion on the individual photon wavepackets. In the presence of dispersion, ``fast'' components of a photon are correlated with the ``slow'' components of its sister photon. For positive dispersion, higher energy (shorter wavelength) components of a light pulse travel faster, while lower energy components lag behind~\cite{Saleh1991}. This leads to a ``chirp''. The minimum and maximum delay ($\tau_{min}$ and $\tau_{max}$) between the detection of the photons determine the spread in observed timing correlations (Figure~\ref{fig:mechanism:a}). For opposite dispersion coefficients (Figure~\ref{fig:mechanism:b}) the chirp imparted on one of the photons is reversed, and the resulting fast-fast/slow-slow correlations minimize the spread in propagation times.

\begin{figure}[b!]
    \centering
    \subfloat{
        \centering
        \includegraphics[width=0.4\linewidth]{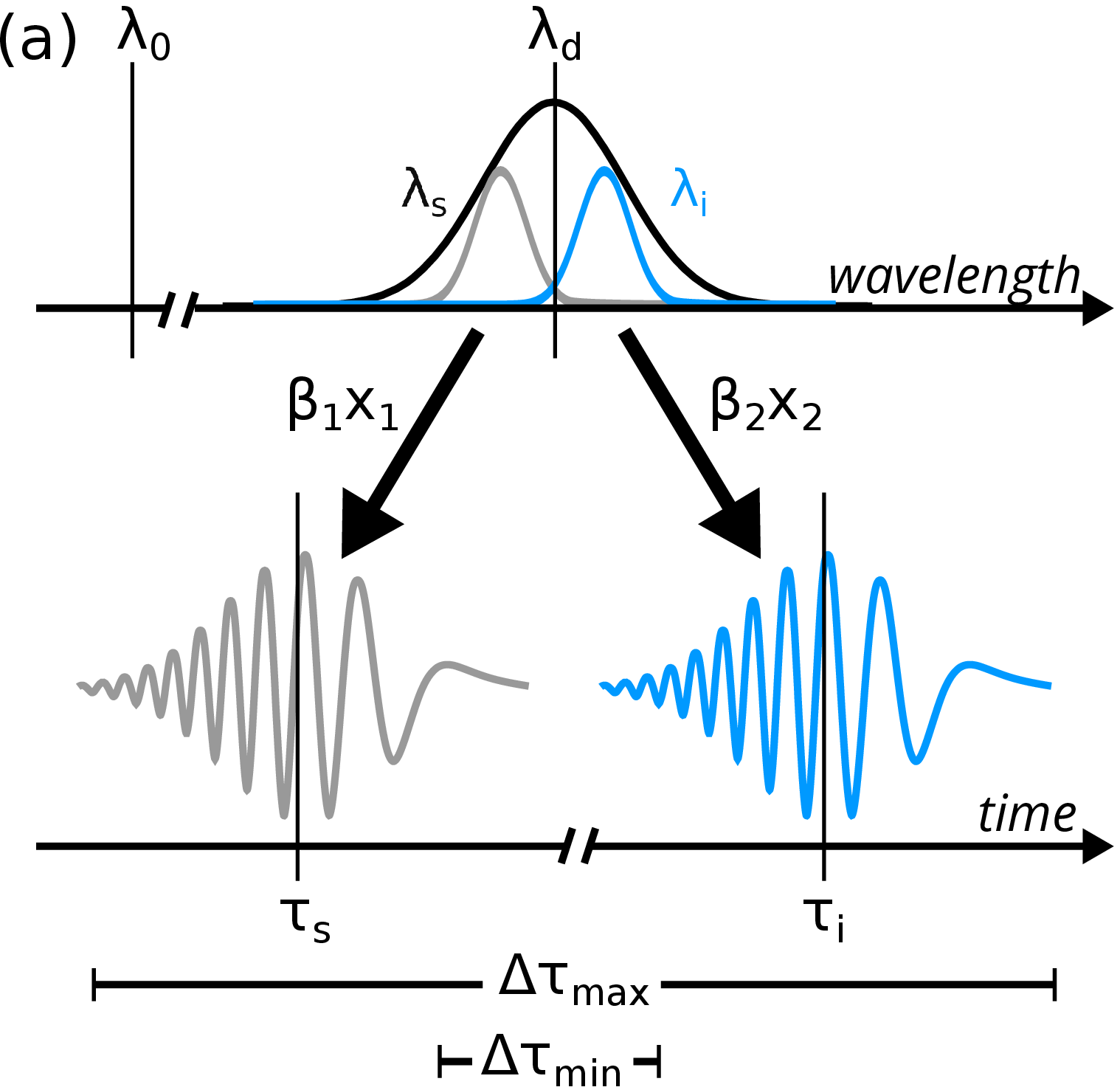}
        \label{fig:mechanism:a}
    }\hspace{1cm}
    \subfloat{
        \includegraphics[width=0.4\linewidth]{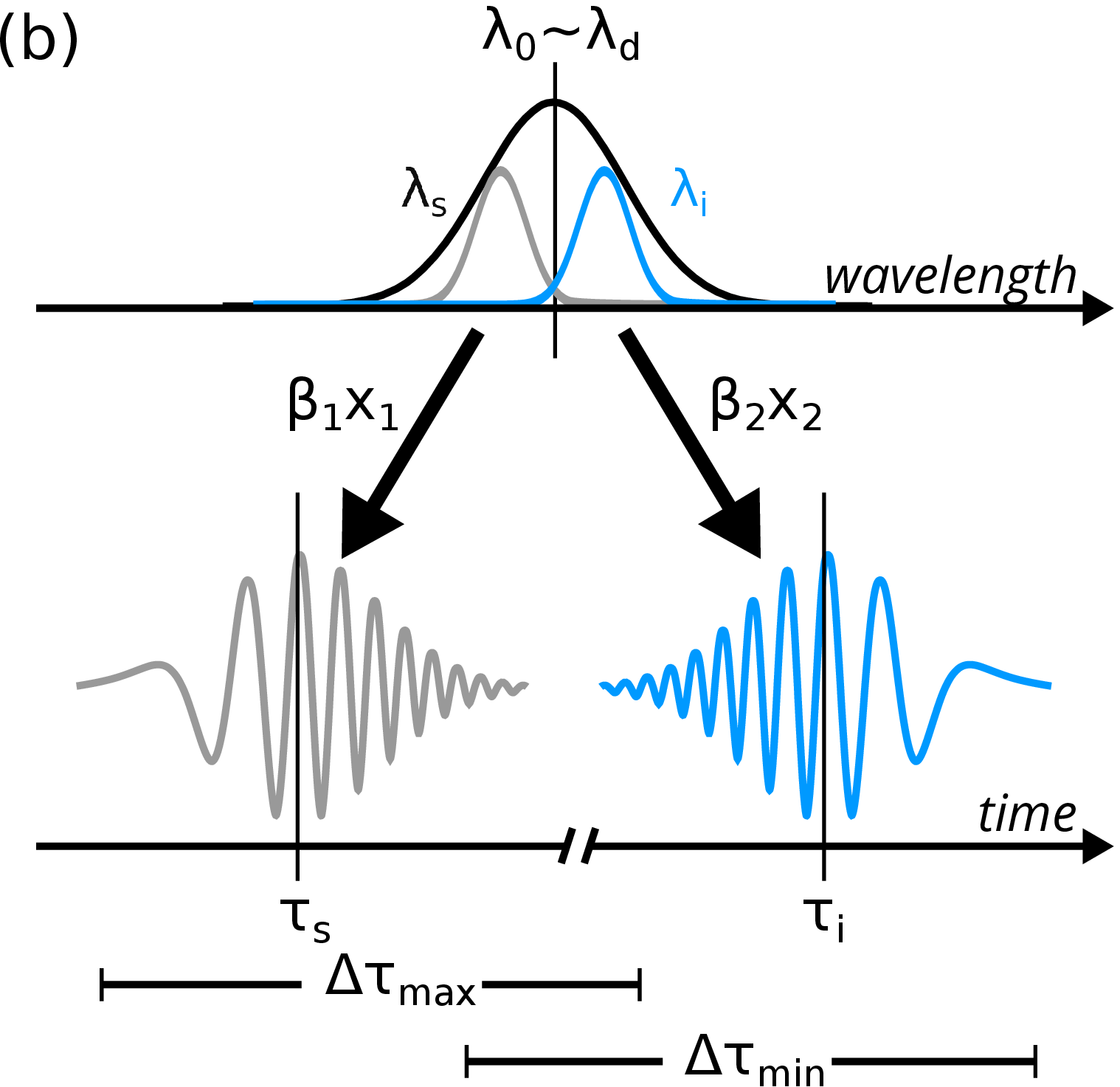}
        \label{fig:mechanism:b}
    }
    \caption{Mechanism of nonlocal dispersion compensation. (a) Time-energy entangled photons are produced about the degenerate wavelength $\lambda_d$, far away from the zero dispersion wavelength $\lambda_0$ of the fiber. Signal and idler photons ($\lambda_s,\lambda_i$) propagate over fibers of length $x_1$ and $x_2$, and are dispersed by $\beta_1x_1$ and $\beta_2x_2$. In this spectral range, the dispersion coefficients $\beta_1$ and $\beta_2$ have the same sign. In the time domain, dispersed photons are represented by their electric field, with their centres denoted $\tau_s$ and $\tau_i$. The width of the pairwise timing correlations will be related to the difference between the minimum and maximum possible delays $\Delta\tau_{min}$ and $\Delta\tau_{max}$ and we note the anticorrelation of the photons in wavelength maximises this discrepancy. (b) Photon pairs with $\lambda_d \sim \lambda_0$ undergo opposite dispersion. In this case, the anticorrelation in wavelength leads to a positive correlation in the observed delay, minimizing $\Delta\tau_{max} - \Delta\tau_{min}$.}
    \label{fig:mechanism}
\end{figure}

%The width of the timing distribution $\sigma$ is related to the sum of the dispersion along the two paths ($\beta_1 x_1$ and $\beta_2 x_2$ for photons 1 and 2, respectively) ~\cite{Franson1992},
%\begin{equation}
%    \sigma^{2} = \frac{ ( \beta_1 x_1 + \beta_2 x_2 )^2 }{2\sigma_0^2},
%    \label{eqn:sigma_franson}
%\end{equation}
%where $\sigma_0$ is the coherence time of the photons, and $x_1$, $x_2$ are the propagation distances. If $\beta_1$ and $\beta_2$ have opposite sign, dispersion can be at least partially compensated. For $\beta_1 x_1 = -\beta_2 x_2$ the compensation is perfect.

This effect has been observed in the visible or near-infrared spectral range by using dispersive elements such as prisms, gratings and specialized fibers~\cite{Steinberg1992,ODonnell2011,Baek2011,Maclean2018}. However, both negative and positive dispersion regions are available in single-mode optical fibers~\cite{Cohen1977}. Most deployed telecommunication fiber exhibits this behaviour around the zero dispersion wavelength close to the \SI{1310}{\nano\meter} ``O-band''~\cite{Ainslie1986}, with the location of this region specified by International Telecommunications Union standards (\emph{ITU-T G.652}~\cite{G652}).

Nonlocal dispersion compensation using the properties of a \emph{single} optical fiber was first observed in measurements of fiber dispersion using SPDC photons~\cite{Brendel1998} and was applied to QKD field tests~\cite{Tittel1998} and entanglement distribution~\cite{Tittel1999}. These experiments utilized a tunable source of SPDC photons to generate wavelengths that would experience dispersion compensation in two continuous spans of deployed telecommunications fiber with lengths up to \SI{9.3}{\kilo\meter}. These early experiments illustrate the potential for nonlocal dispersion compensation to increase the signal-to-noise ratio of a quantum channel.

In this paper, we show that photon pairs broadly degenerate at the approximate location of the zero dispersion wavelength can exhibit nonlocal dispersion compensation in standard, multi-segment telecommunication fiber. This scheme does not require specialized dispersive elements, measurement of the precise fiber characteristics or tuning of the emission spectrum.

Nonlocal dispersion compensation can be understood by considering the energy anticorrelation of an entangled photon pair and dispersion on the individual photon wavepackets. In the presence of dispersion, ``fast'' components of a photon are correlated with the ``slow'' components of its sister photon. For positive dispersion, higher energy (shorter wavelength) components of a light pulse travel faster, while lower energy components lag behind~\cite{Saleh1991}. This leads to a ``chirp''. The minimum and maximum delay ($\tau_{min}$ and $\tau_{max}$) between the detection of the photons determine the spread in observed timing correlations (Figure~\ref{fig:mechanism}a). For opposite dispersion coefficients (Figure~\ref{fig:mechanism}b) the chirp imparted on one of the photons is reversed, and the resulting fast-fast/slow-slow correlations minimize the spread in propagation times.

The width $\sigma$ of the timing distribution is related to the sum of the dispersion along the two paths ($\beta_1 x_1$ and $\beta_2 x_2$ for photons 1 and 2, respectively) ~\cite{Franson1992},
\begin{equation}
    \sigma^{2} = \frac{ ( \beta_1 x_1 + \beta_2 x_2 )^2 }{2\sigma_0^2},
    \label{eqn:sigma_franson}
\end{equation}
where $\sigma_0$ is the coherence time of the photons, and $x_1$, $x_2$ are the propagation distances. If the dispersion coefficients $\beta_1$ and $\beta_2$ have opposite sign, dispersion can be at least partially compensated. For $\beta_1 x_1 = -\beta_2 x_2$ the compensation is perfect~\cite{Franson1992}.

\begin{figure}[t]
    \centering
    \includegraphics[width=0.8\linewidth]{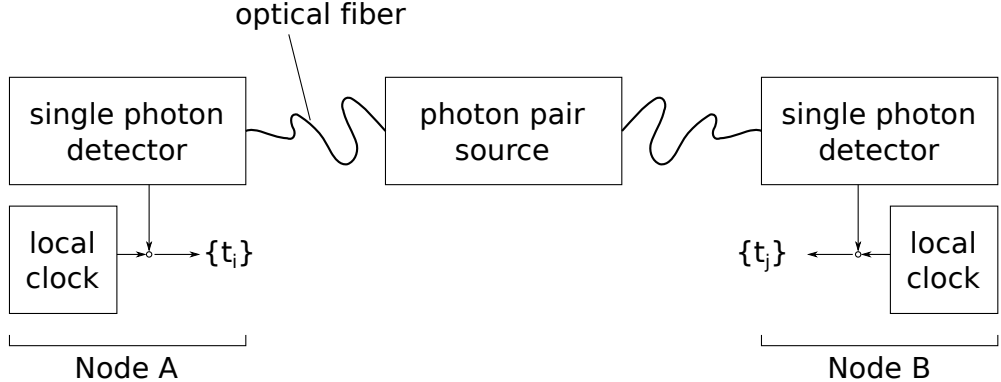}
    \caption{A schematic of the experimental setup. Time correlated photon pairs are generated by Type-0 spontaneous parametric downconversion. The pairs are separated, routed to two nodes via single mode optical fiber, and detected by InGaAs avalanche photodiodes. Detection times are recorded with respect to a local clock.}
    \label{fig:system}
\end{figure}

Figure~\ref{fig:system} shows a schematic of the experimental setup. A photon pair source is connected to two remote nodes by optical fiber. At the nodes, arrival times of single photons are recorded with respect to a local clock. Due to the timing correlation of the photon pairs, detection times $\{t_i\}$ and $\{t_j\}$ at node A and B are correlated. As photon pairs are created at random time intervals, their detection results in a random set of arrival times at each node. These are processed into sets of delays ($d_a$ and $d_b$) such that

\begin{equation}
    d_a(t) = \sum_i \delta (t - t_i);~~~~
    d_b(t) = \sum_j \delta (t - t_j),
    \label{eqn:d}
\end{equation}

\noindent with their cross correlation c($\tau$)

\begin{equation}
    c(\tau) = \int d_a(t ) d_b (t + \tau) dt.
    \label{eqn:ct}
\end{equation}

This cross correlation will exhibit a peak at a delay $\tau$ corresponding to the relative time of flight of the single photons. The identification of this peak allows the pairing of corresponding detection events~\cite{Marcikic2006}. In several entanglement-based quantum key distribution systems~\cite{Marcikic2006,Ursin2007}, the presence of significant propagation losses along with a degree of background noise makes minimizing the width of this peak an important consideration~\cite{Diamanti2016}.

% Problems with the gradients on lines in this figure. Not sure how to resolve.
%\begin{figure}
%    \centering
%    \includegraphics[width=0.8\linewidth]{dispersion_compensation.pdf}
%   \caption{A schematic illustration of nonlocal dispersion compensation. Dispersed photons are represented by Gaussian envelopes representing their distribution in time, with their spectral characteristics depicted using a blue-red color palette. The width of the resulting distribution of coincident detction events is depicted as the combination of minimum and maximum delays ($t_{long}, t_{short}$).  (a) In the case of only one photon being dispersed, the minimum delay between coincident detection events is given by the delay between the detection of the undispersed photon and the leading edge of the dispersed photon. (b) Where both photons experience dispersion with the same sign, the difference between minimum and maximum delays is maximised due to the anticorrelation of photon pairs in energy. (c) For the case where dispersion of the photons have opposite sign, this anticorrelation gives rise to correlated transit times, reducing the difference between minimum and maximum delays.}
%    \label{fig:my_label}
%\end{figure}

\begin{figure}[ht!]
    \centering
    \includegraphics[width=0.7\linewidth]{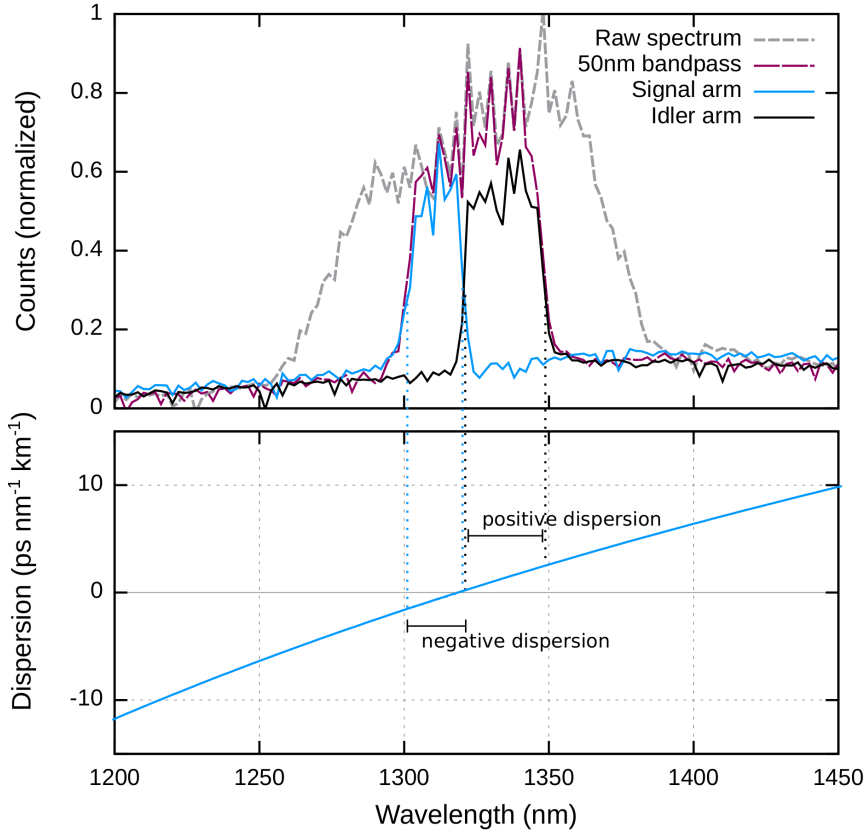}
    \caption{Spectrum of correlated photons in this experiment. Photon pairs are degenerate around \SI{1316}{\nano\meter}. In a \SI{50}{\nano\meter} window defined by a bandpass filter, signal and idler photons are separated by a wavelength division demultiplexer with an edge at approximately \SI{1316}{\nano\meter}. The bottom panel shows the dispersion of the popular SMF-28e (Corning~\cite{SMF28e}), a fiber conforming to the ITU-T G.652 standard~\cite{G652}. Dispersion is shown for a zero dispersion wavelength ($\lambda_0$) of \SI{1316}{\nano\meter}, compatible with the specified range (\SIrange{1302}{1322}{\nano\meter}). Signal and idler photons propagating in this fiber will experience negative and positive dispersion.}
    \label{fig:spectrum}
\end{figure}

Our photon pair source is based on Type-0 SPDC in a periodically poled crystal of potassium titanyl phosphate (PPKTP, Raicol) pumped by a grating stabilized laser diode at \SI{658}{\nano\meter} (Ondax). The resulting photon pairs are degenerate at \SI{1316}{\nano\meter}, close to the zero dispersion wavelength in the most common single-mode telecommunications fibers~\cite{G652}, with emission sufficiently broad to span a region on either side of this wavelength (see Figure~\ref{fig:spectrum}). Signal and idler photons are efficiently separated using a wavelength division demultiplexer, and routed to either a deployed fiber link or a bank of lab-based fibers. After propagation and dispersion, we detect the photons using commercially available InGaAs avalanche photodiodes (APDs) operated in Geiger mode and record arrival times using timestamping modules.

There is an intrinsic uncertainty in the delay between the detection of a photon and the emission of a macroscopic electrical signal. For avalanche photodiodes, this jitter is usually of the order of hundreds of picoseconds~\cite{Lunghi2012,Stipcevic2013}, while for superconducting nanowire sensors it can be as little as tens of picoseconds~\cite{Natarajan2012}. This uncertainty, along with the resolution of timestamping electronics provides a lower bound to the width of correlation $c(\tau)$. The InGaAs APDs used in our work exhibit a jitter of \SI{87}{\pico\second} and \SI{110}{\pico\second}.

\begin{figure}[h!]
    \centering
    \includegraphics[width=0.7\linewidth]{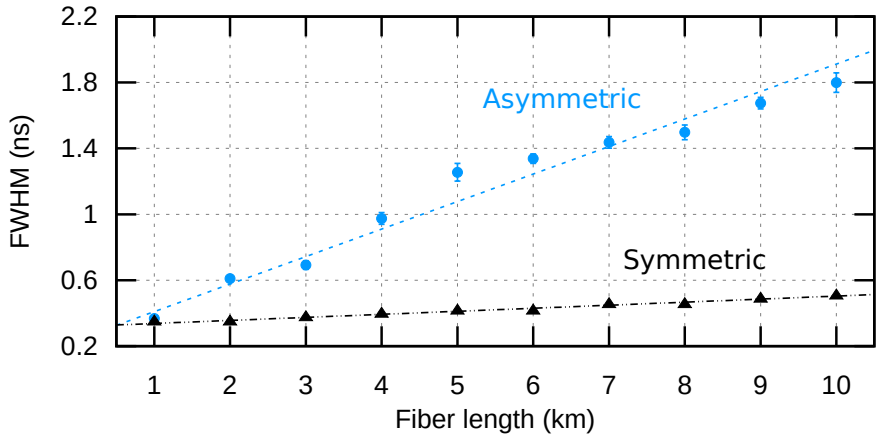}
    \caption{Autocorrelation width (FWHM) for photon pairs after propagation through various lengths of SMF-28e optical fiber. We measure the asymmetric case of one photon detected immediately (after negligible dispersion) while the other photon is transmitted down the fiber, with the FWHM exhibiting an increase of \SI{167}{\pico\second\per\kilo\meter}. Also shown is the symmetric case of both photons propagating in the same fiber. Here tne trend is much reduced, with a linear increase in FWHM of \SI{18}{\pico\second\per\kilo\meter}.}
    \label{fig:fwhm}
\end{figure}

To probe the interaction of photon pairs with the dispersive properties of optical fiber, we transmit photons through several lengths of fiber from \SIrange{1}{10}{\kilo\meter}, cut from the same piece in order to maintain similar zero dispersion wavelengths. Figure~\ref{fig:fwhm} shows the width of $c(\tau)$ for the assymetric case of one photon detected directly while the other is first dispersed by an optical fiber. An approximately linear relationship is observed between propagation distance and correlation width, with gradient \SI{167}{\pico\second\per\kilo\meter}. We also investigate the symmetric case where both photons are transmitted over the same fiber, before being separated and detected. In the symmetric case, dispersion is reduced to \SI{18}{\pico\second\per\kilo\meter}. This reduction is in agreement with Equation~\ref{eqn:sigma_franson}, consistent with $\beta_1 \sim -\beta_2$. While perfect compensation could be achieved by tailoring the degenerate wavelength $\lambda_d$ to $\lambda_0$ of the specific fiber, this is impractical in deployed networks comprising fibers with different $\lambda_0$.

\begin{figure}[h!]
    \centering
    \includegraphics[width=0.8\linewidth]{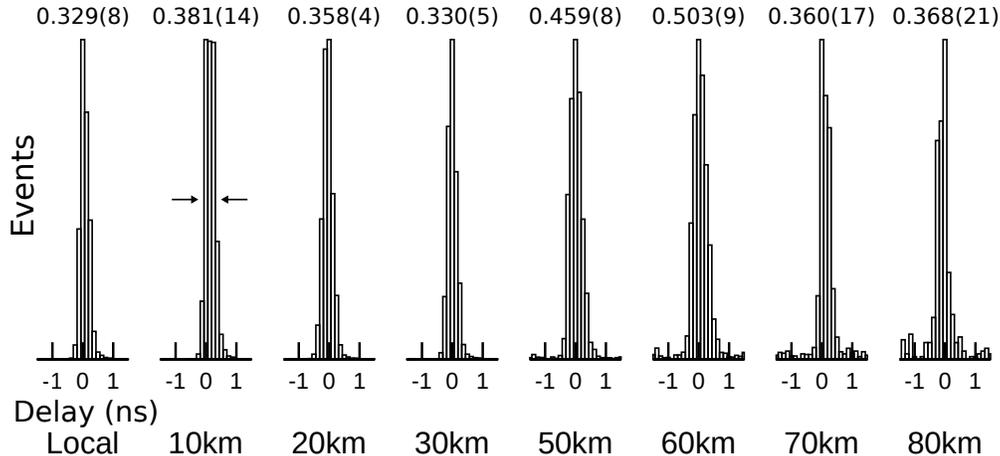}
    \caption{Timing correlations for photon pairs transmitted through multi-segment optical fibers in a laboratory environment. The horizontal axis resolution is limited by the timestamping electronics to \SI{125}{\pico\second}, with fitted FWHM values indicated above each graph. In the minimally dispersed case (``local''), photons are routed to detectors via approximately \SI{4}{\meter} of fiber. We observe the preservation of a high degree of timing correlation, despite transmission through up to \SI{80}{\kilo\meter} fiber.}
    \label{fig:histograms}
\end{figure}

We carry out the symmetric measurement for longer fibers, with correlation signals shown in Figure~\ref{fig:histograms}. These fibers are comprised of several segments connected in series, with the longest (\SI{80}{\kilo\meter}) made up of three segments (10, 20, \SI{50}{\kilo\meter}). We no longer observe the linear increase of dispersion with fiber length (Figure~\ref{fig:fwhm}). However, tight timing correlations are preserved ($<$\SI{0.503(9)}{\nano\second}). We attribute small differences in the degree of dispersion compensation to variation in the exact position of the zero dispersion wavelength, which by specification may lie in a relatively wide range of \SIrange{1302}{1322}{\nano\meter}~\cite{SMF28e}.

It is interesting to note that the degree of compensation seen in the series of shorter fibers is lower than for any of the longer spools. For example the observed FWHM after \SI{10}{\kilo\meter} of symmetric propagation is \SI{0.506(7)}{\nano\second}, compared with \SI{0.381(14)}{\nano\second} (Figure~\ref{fig:histograms}). This observation suggests that significant compensation is possible without tuning $\lambda_d$ of the source.

\begin{figure}[ht!]
    \centering
    \subfloat{
        \centering
        \includegraphics[width=0.8\linewidth]{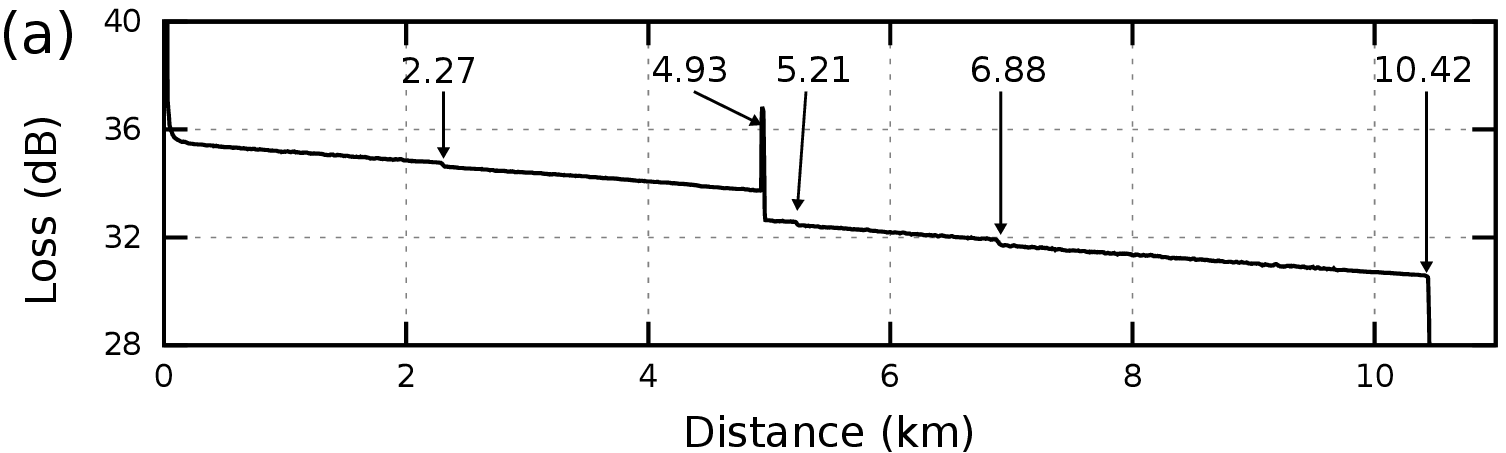}
        \label{fig:10km:otdr}
    }\\
    
    \subfloat{
        \centering
        \includegraphics[width=0.4\linewidth]{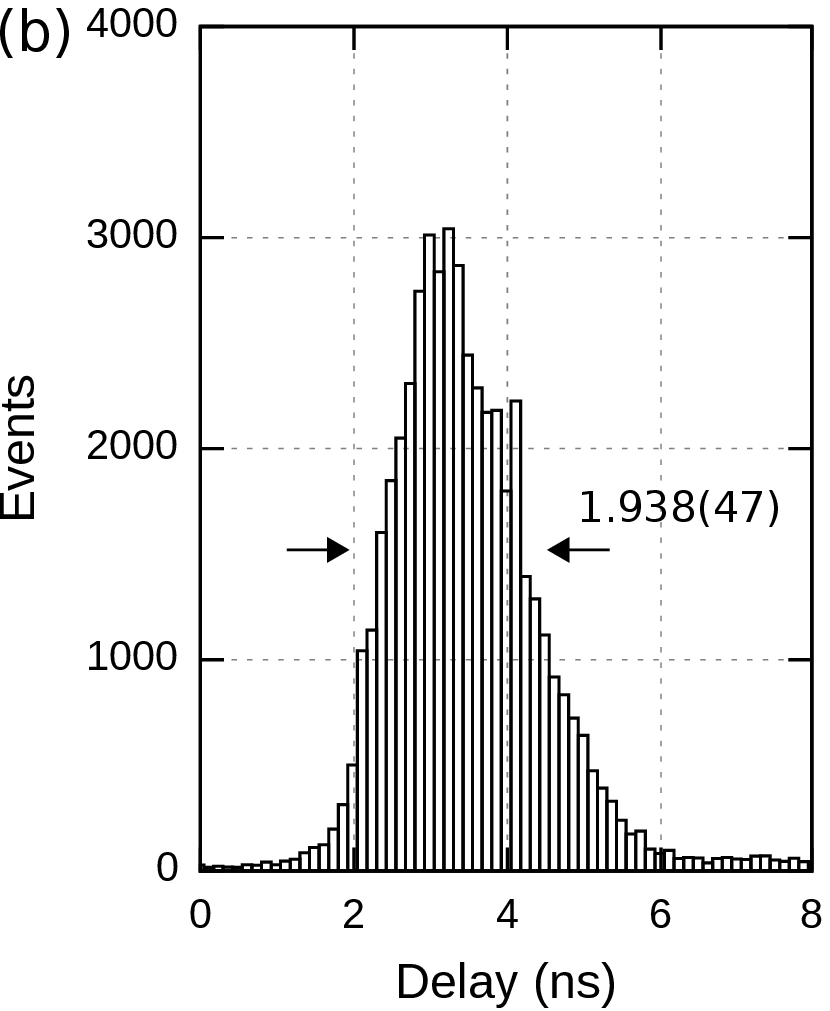}
        %\caption{$x_a$ = \SI{0}{\kilo\meter}, $x_b$ = \SI{10}{\kilo\meter}}
        \label{fig:10km:a}
    }
    \subfloat{
        \includegraphics[width=0.4\linewidth]{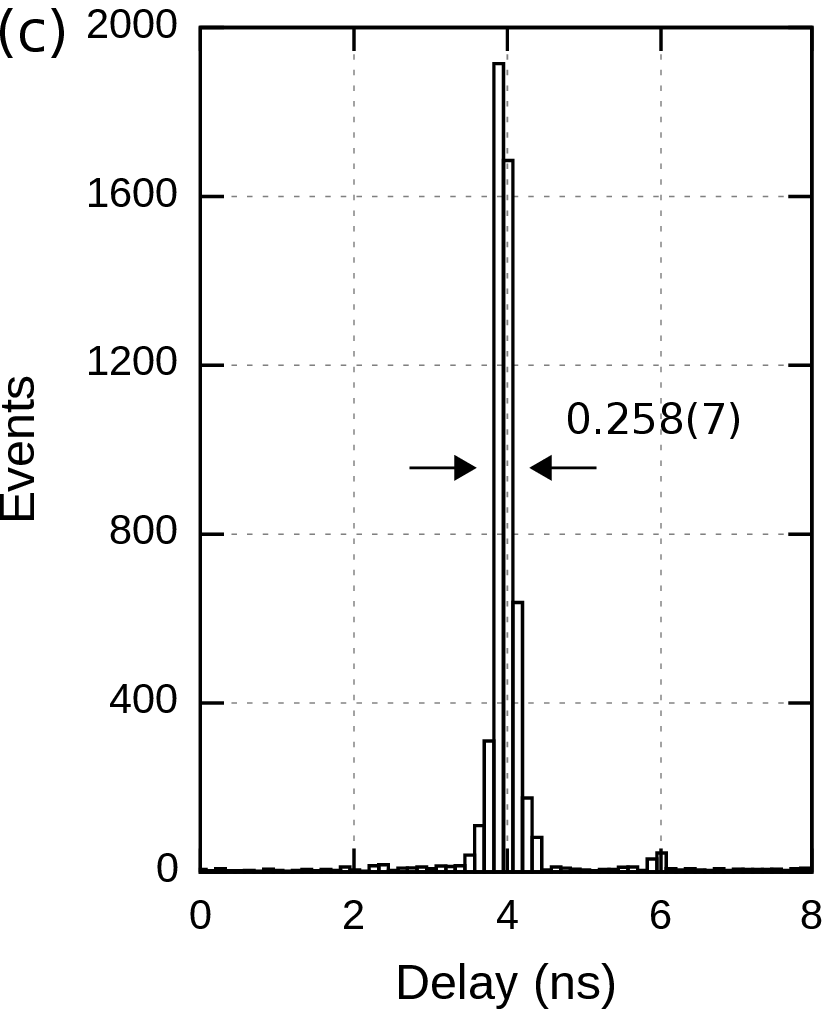}
        %\caption{$x_a$ = \SI{10}{\kilo\meter}, $x_b$ = \SI{10}{\kilo\meter}}
        \label{fig:10km:b}
    }
    \caption{Measurements of photon pairs propagating in deployed telecommunication fiber. (a) \SI{1310}{\nano\meter} OTDR measurement of one fiber span, showing a length of \SI{10.42}{\kilo\meter}. At least four splices are observed. (a,b) Cross correlation histograms, (b) one photon is detected locally while the other propagates over a deployed fiber prior to detection; (c) both photons travel through separate deployed fibers. Fiber spans formed two loops, beginning and ending in the lab. Photons were detected using InGaAs avalanche photodiodes, and timestamped at a nominal resolution of \SI{125}{\pico\second} (the small feature observed at \SI{4}{\nano\second} is an artifact of the data acquisition). Delays of \SI{52230}{\nano\second} and \SI{128}{\nano\second} have been subtracted from (a) and (b) respectively for ease of comparison.}
    \label{fig:10km}
\end{figure}

To test this mechanism in an operationally useful context we transmit photons through two separate \SI{10}{\kilo\meter} spans of deployed telecommunication fiber. An optical time domain reflectometer (OTDR) measurement for one fiber is shown in Figure~\ref{fig:10km:otdr}, revealing at least five segments. From our previous observations we do not expect these segments to exhibit identical zero dispersion wavelengths. Measured $c(\tau)$ histograms for one photon transmitted and one detected locally and for both photons transmitted are shown in Figure~\ref{fig:10km:a} and \ref{fig:10km:b}. With only one photon transmitted, chromatic dispersion results in a coincidence distribution with a FWHM of \SI{1.938(47)}{\nano\second}. When both photons are transmitted over separate fibers, we observe a distribution with FWHM \SI{0.258(7)}{\nano\second}.

Laboratory and field test measurements unambiguously demonstrate that photon pairs with appropriately engineered spectral properties can experience self-compensation of dispersion in conventional telecommunication fiber networks. This is despite the presence of a range of zero dispersion wavelengths and accomplished without the requirement of source tuning. This capability paves the way for the use of broad spectrum entangled light sources for quantum key distribution and other forms of quantum communication. The use of the intrinsic anomalous dispersion available in standard telecommunications fiber can minimize or even remove the need for specialized dispersion-compensating apparatus. The trade-off of operating in the O-band (where attenuation losses are higher than in the more commonly used C-band) will be acceptable for many use cases, particularly for metropolitan areas with substantial existing fiber infrastructure.

\section*{Funding}
This research is supported by the National Research Foundation, Prime
Minister’s Office, Singapore under its Corporate Laboratory@University
Scheme, National University of Singapore, and Singapore
Telecommunications Ltd.

\section*{Acknowledgements}
The authors thank Amelia Tan Peiyu and the Singtel fiber team for facilitating our deployed fiber tests.

% Bibliography
\bibliography{spdc-compensation}

% Full bibliography will be added automatically on a new page for Optics Letters submissions. This command is ignored for journal article submissions.
% Note that this extra page will not count against page length.
%\bibliographyfullrefs{spdc-compensation}

%Manual citation list
%\begin{thebibliography}{1}
%\bibitem{Zhang:14}
%Y.~Zhang, S.~Qiao, L.~Sun, Q.~W. Shi, W.~Huang, %L.~Li, and Z.~Yang,
 % \enquote{Photoinduced active terahertz metamaterials with nanostructured
  %vanadium dioxide film deposited by sol-gel method,} Opt. Express \textbf{22},
  %11070--11078 (2014).
%\end{thebibliography}

\end{document}